\documentstyle[12pt,epsf]{article}
\def\bm#1{\mbox{\boldmath $#1$}}
\def\s{\sigma}
\def\I{\rm i}
\newcommand{\be}{\begin{eqnarray}}
\newcommand{\ee}{\end{eqnarray}}
\newcommand{\no}{\nonumber}
\def\case#1#2{\textstyle{#1\over#2}\displaystyle}

\begin{document}
\begin{center}
{\bf \LARGE Exactly solvable quantum spin tubes\\ and ladders\\}
\vspace{8mm}
{\bf \large M T Batchelor and M Maslen\\}
\vspace{2mm}
{Department of Mathematics, School of Mathematical Sciences,\\
Australian National University, Canberra ACT 0200, Australia\\}
\end{center}
\vspace{6mm}
\begin{abstract}
We find families of integrable $n$-leg spin-$\case{1}{2}$ ladders and
tubes with general isotropic exchange interactions between spins.
These models are equivalent to $su(N)$ spin chains with $N=2^n$.
Arbitrary rung interactions in the spin tubes and ladders
induce chemical potentials in the equivalent spin chains.
The potentials are $n$-dependent and differ for the tube and ladder models.
The models are solvable by means of nested Bethe Ansatz.
\end{abstract}
\vspace{6mm}

The physics of quantum spin ladders has attracted a great deal of
recent theoretical and experimental interest. 
It is now well established that the properties of $n$-leg Heisenberg 
spin-$\case12$ ladders show a remarkable $n$-dependence \cite{DR}. 
For $n$ odd the Heisenberg ladders have a gapless ground state with 
a quasi long-range order, while for $n$ even there is a spin liquid 
ground state with short-range correlations and an energy gap. 
A number of different compounds have been found which confirm this behaviour. 

Of course, for $n=1$ these properties are borne out by the solvable 
Heisenberg chain.
However, the more general $n$-leg Heisenberg ladders defy an exact solution. 
Nevertheless some solvable spin ladders have been found. 
In recent progress Wang \cite{W} discussed a 2-leg ladder with hamiltonian   
\begin{eqnarray}
H&=&\case{1}{4} \sum_{j=1}^L \left( \bm{\s}_j^{(1)} \cdot \bm{\s}_{j+1}^{(1)}
+ \bm{\s}_j^{(2)} \cdot \bm{\s}_{j+1}^{(2)} \right)
+\case{1}{2} J \sum_{j=1}^L \bm{\s}_j^{(1)} \cdot \bm{\s}_j^{(2)} \no\\
&\phantom{=}& +\case{1}{4} \sum_{j=1}^L \left( \bm{\s}_j^{(1)} \cdot
\bm{\s}_{j+1}^{(1)} \right)
\left( \bm{\s}_j^{(2)} \cdot \bm{\s}_{j+1}^{(2)} \right)
\end{eqnarray}
where $\bm{\s}_j^{(1)}$ and $\bm{\s}_j^{(2)}$ are Pauli matrices acting
at site $j$ on legs 1 and 2 of the ladder, with  
$\bm{\s}_j = (\s_j^x, \s_j^y, \s_j^z)$.
$L$ is the number of rungs and periodic boundary conditions are imposed.
It is to be noted that the last term of the hamiltonian defines biquadratic 
interactions which couple sites on the two legs of the ladder. 
Such interchain coupling can be of experimental importance \cite{W}. 
The variable parameter $J$ measures the strength of the rung interactions.
Another, more complicated, though with no variable parameter, 
solvable 2-leg hamiltonian has been introduced by
Albeverio and Fei \cite{AF}, about which we will say more elsewhere.
Yet other 2-leg hamiltonians have been defined with special matrix-product
groundstates \cite{KM}.
Our interest here is in integrable or exactly solvable models with an 
underlying $R$-matrix.

The solvability of Wang's model lies in the observation that it can be
mapped to the hamiltonian of an $su(4)$-invariant spin chain for $J=0$,
whilst for $J\ne0$ the rung interactions take the form of a chemical potential. 
Specifically, hamiltonian (1) can be written \cite{W} 
\be
H&=&\case{1}{4} \sum_{j=1}^L
\left( 1 + \bm{\s}_j^{(1)} \cdot \bm{\s}_{j+1}^{(1)} \right)
\left( 1 + \bm{\s}_j^{(2)} \cdot \bm{\s}_{j+1}^{(2)} \right) \no\\
&\phantom{=}& +\case{1}{2} J \sum_{j=1}^L 
\left( \bm{\s}_j^{(1)} \cdot \bm{\s}_j^{(2)} - 1 \right) 
+ \case{1}{2} (J - \case{1}{2}) L \no\\
&=&\sum_{j=1}^L \sum_{\alpha,\beta=1}^4 
X_j^{\alpha\beta}X_{j+1}^{\beta\alpha}-2J\sum_{j=1}^L X_j^{11}
\ee
up to an irrelevant constant.
The operators $X_j^{\alpha\beta} = | \alpha_j \rangle \langle \beta_j |$,
where $| \alpha_j \rangle$ are the (orthogonalised) eigenstates of the
single rung hamiltonian.
The constant $2J$ in the last term indicates a chemical potential applied 
on $N_1$, where in general the operators 
$N_\alpha=\sum_{j=1}^N X_j^{\alpha\alpha}$ are conserved quantities. 

The underlying integrability of the 2-leg ladder hamiltonian (1) is thus
seen to be due to the known $R$-matrix associated with $su(4)$ \cite{BY}. 
The key ingredient is the permutation operator 
\be
P_{j,j+1} = \case{1}{4} 
\left( 1 + \bm{\s}_j^{(1)} \cdot \bm{\s}_{j+1}^{(1)} \right)
\left( 1 + \bm{\s}_j^{(2)} \cdot \bm{\s}_{j+1}^{(2)} \right)
= \sum_{\alpha,\beta=1}^4 X_j^{\alpha\beta}X_{j+1}^{\beta\alpha}
\ee  
Our starting point is to note that for an $n$-leg ladder the permutation
operator can be written 
\be
P_{j,j+1} = \frac{1}{N} \prod_{i=1}^n 
\left( 1 + \bm{\s}_j^{(i)} \cdot \bm{\s}_{j+1}^{(i)} \right)
= \sum_{\alpha,\beta=1}^N X_j^{\alpha\beta}X_{j+1}^{\beta\alpha}
\ee
in which $X_j^{\alpha\beta}$ are $N \times N$ $su(N)$ operators 
with $N = 2^n$. 
Thus we can immediately write down a family of solvable $n$-leg ladders
with hamiltonian
\begin{eqnarray}
H_{n}^{\rm ladder}= \case{1}{2^n} \sum_{j=1}^L \prod_{i=1}^n
\left( 1 + \bm{\s}_j^{(i)} \cdot \bm{\s}_{j+1}^{(i)} \right)
\label{n-ladder}
\end{eqnarray}
The operators $\bm{\s}_j^{(i)}$ are defined on leg $i$. 
The 3-leg hamiltonian reads
\begin{eqnarray}
H_{3}^{\rm ladder}&=& \case{1}{8} \sum_{j=1}^L \left[ 
\bm{\s}_j^{(1)} \cdot \bm{\s}_{j+1}^{(1)} +
\bm{\s}_j^{(2)} \cdot \bm{\s}_{j+1}^{(2)} +
\bm{\s}_j^{(3)} \cdot \bm{\s}_{j+1}^{(3)} 
 \right. \no
\\
&\phantom{=}& \left. + 
\left( \bm{\s}_j^{(1)} \cdot \bm{\s}_{j+1}^{(1)} \right) 
\left( \bm{\s}_j^{(2)} \cdot \bm{\s}_{j+1}^{(2)} \right) + 
\left( \bm{\s}_j^{(1)} \cdot \bm{\s}_{j+1}^{(1)} \right) 
\left( \bm{\s}_j^{(3)} \cdot \bm{\s}_{j+1}^{(3)} \right)
\right. \no\\
&\phantom{=}& \left. + 
\left( \bm{\s}_j^{(2)} \cdot \bm{\s}_{j+1}^{(2)} \right)
\left( \bm{\s}_j^{(3)} \cdot \bm{\s}_{j+1}^{(3)} \right)
\right. \no\\
&\phantom{=}& \left. + 
\left( \bm{\s}_j^{(1)} \cdot \bm{\s}_{j+1}^{(1)} \right)
\left( \bm{\s}_j^{(2)} \cdot \bm{\s}_{j+1}^{(2)} \right) 
\left( \bm{\s}_j^{(3)} \cdot \bm{\s}_{j+1}^{(3)} \right)
\right]
+ \case{1}{8} L
\end{eqnarray}
In general this family of $n$-leg ladders includes up to 
$n$-body interactions. As the number of legs increases these
interactions become increasingly nonlocal, with for example,
$\bm{\s}_j^{(1)} \cdot \bm{\s}_{j+1}^{(1)}$ on leg 1 interacting with 
$\bm{\s}_j^{(n)} \cdot \bm{\s}_{j+1}^{(n)}$ on leg $n$.
However, such interactions are necessary to preserve the integrability 
of the model.\footnote{The situation is somewhat akin to the price
paid in the integrable spin-$S$ generalisations of the Heisenberg chain
for which terms up to order $(\bm{S}_j \cdot \bm{S}_{j+1})^{2S}$
appear \cite{BY}.}

To some extent, the nonlocality can be overcome by considering 
quantum spin ``tubes'', rather than spin ladders. The 3- and 4-tubes
are depicted in Figure 1. 
The $n$-leg ladder hamiltonian (\ref{n-ladder}) applies equally well 
to the $n$-tube.
In this way the above hamiltonian $H_{3}^{\rm ladder}$ is equal
to $H_{3}^{\rm tube}$.
The eigenspectra of the hamiltonians $H_{n}^{\rm ladder}$ and
$H_{n}^{\rm tube}$ are equivalent to that of the $su(N)$ 
permutation hamiltonian \cite{S}, with eigenvalues
\be
E=L-\sum_{j=1}^{M_1}\frac 1{\left(\lambda_j^{(1)}\right)^2+\frac14}
\ee
where we recall that $N=2^n$.
The $su(N)$ Bethe equations are given in terms of $N-1$ roots 
$\lambda_j^{(r)}$
\begin{eqnarray}
\qquad
\left(\frac{\lambda_j^{(1)}-\case{\I}{2}}
           {\lambda_j^{(1)}+\case{\I}{2}} \right)^L 
&=&\prod_{k\neq j}^{M_1} \frac{\lambda_j^{(1)}-\lambda_k^{(1)} -\I}
                            {\lambda_j^{(1)}-\lambda_k^{(1)} +\I}
\prod_{k=1}^{M_2}      \frac{\lambda_j^{(1)}-\lambda_k^{(2)}+\case{\I}{2}}
                            {\lambda_j^{(1)}-\lambda_k^{(2)}-\case{\I}{2}},
\nonumber\\
\prod_{k\neq j}^{M_r} \frac{\lambda_j^{(r)}-\lambda_k^{(r)}-\I}
                           {\lambda_j^{(r)}-\lambda_k^{(r)}+\I}
&=&\prod_{k=1}^{M_{r-1}} \frac{\lambda_j^{(r)}-\lambda_k^{(r-1)}-\case{\I}{2}}
                            {\lambda_j^{(r)}-\lambda_k^{(r-1)}+\case{\I}{2}}
 \prod_{k=1}^{M_{r+1}} \frac{\lambda_j^{(r)}-\lambda_k^{(r+1)}-\case{\I}{2}}
                            {\lambda_j^{(r)}-\lambda_k^{(r+1)}+\case{\I}{2}}
\label{nba}
\end{eqnarray}
where $j = 1, \ldots, M_r$ with $r = 1, \ldots, N-1$. We take $M_N = 0$.
It is well-known that the isotropic $su(N)$ models are critical with no gap.

Note that the permutation operator for the $su(N)$ models can be
written in terms of spin-$S$ operators as
\be
P_{j,j+1} = (-)^{2S} \sum_{i=0}^{2S} \prod_{k\ne i}^{2S}
\frac{Y_j - x_k}{x_i - x_k}
\ee
where $Y_j = \bm{S}_j \cdot \bm{S}_{j+1}$,    
$x_k = \case{1}{2} k(k+1) - S(S+1)$ and $N = 2S +1$.
This gives an equivalent, $su(2)$-invariant, representation of the 
eigenspectrum.
The hamiltonian of the $n$-leg spin ladder or spin tube is of size
$2^{n L} \times 2^{n L}$, which is equivalent to that of the $su(N)$
chain of length $L$, namely $N^L \times N^L$.

\begin{figure}
\begin{center}
\hspace{5mm}
\epsfxsize=3.0in
\epsfbox{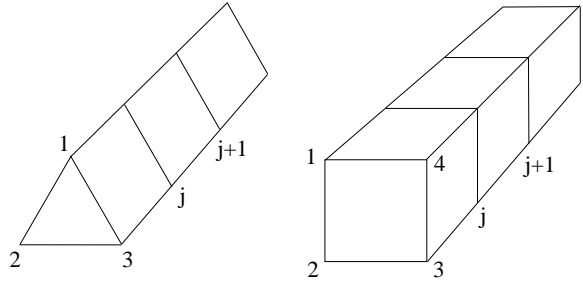}
\end{center}
\caption{Labelling of legs and rungs on the 3- and 4-leg spin tubes.}
\end{figure}

Rung interactions of variable strength can also be introduced in these
models.
For the ladder and tube hamiltonians, we define
\begin{eqnarray}
H_{n}^{\rm ladder}(J)&=& \sum_{j=1}^L \left[ \case{1}{2^n} \prod_{i=1}^n 
\left( 1 + \bm{\s}_j^{(i)} \cdot \bm{\s}_{j+1}^{(i)} \right)
+\case{1}{2} J \sum_{i=1}^{n-1} 
\left( \bm{\s}_j^{(i)} \cdot \bm{\s}_j^{(i+1)} - 1 \right) \right] \,\,\, 
\label{hlad}
\end{eqnarray}
\begin{eqnarray}
H_{n}^{\rm tube}(J)&=& \sum_{j=1}^L \left[ \case{1}{2^n} \prod_{i=1}^n 
\left( 1 + \bm{\s}_j^{(i)} \cdot \bm{\s}_{j+1}^{(i)} \right)
+\case{1}{2} J \sum_{i=1}^{n} 
\left( \bm{\s}_j^{(i)} \cdot \bm{\s}_j^{(i+1)} - 1 \right) \right]   
\label{htub}
\end{eqnarray}
where for the tube we have $\bm{\s}_j^{(n+1)}=\bm{\s}_j^{(1)}$.
The equivalent hamiltonians can be contructed from a consideration
of the rung basis states. 
The first term in Eqs. (\ref{hlad}) and (\ref{htub}) are simply 
permutation operators.
The second terms can also be expressed in terms of $X$ operators,
using the fact that any hamiltonian $h$ may be expressed as 
$h = \sum_i \lambda_i | \alpha_i \rangle \langle \alpha_i | = \sum_i
\lambda_i X^{ii}$.
Here $\lambda_i$ is the eigenvalue corresponding to eigenstate
$| \alpha_i \rangle$ and the sum is over all eigenstates.
We choose to write
$h = \sum_i (\lambda_i - \lambda_{\rm max}) X^{ii}$,
where $\lambda_{\rm max}$ is the largest eigenvalue.
This adds a constant to the hamiltonian but does not change the underlying
physics.
It has the advantage of leading to a more compact representation of $h$
in terms of $X$ matrices.
We list here the first few cases:
\be
H_{3}^{\rm ladder}(J)&=& \sum_{j=1}^L\sum_{\alpha,\beta=1}^8
X_j^{\alpha\beta}X_{j+1}^{\beta\alpha}-J\sum_{j=1}^L \left(
X_j^{11}+3X_j^{22}+3X_j^{33}+X_j^{44} \right) \,\,\,
\ee
\be
H_{3}^{\rm tube}(J)&=& \sum_{j=1}^L\sum_{\alpha,\beta=1}^8
X_j^{\alpha\beta}X_{j+1}^{\beta\alpha}-3J\sum_{j=1}^L \left(
X_j^{11}+X_j^{22}+X_j^{33}+X_j^{44} \right)
\ee
\be
H_{4}^{\rm tube}(J)&=& \sum_{j=1}^L\sum_{\alpha,\beta=1}^{16}
X_j^{\alpha\beta}X_{j+1}^{\beta\alpha}-2J\sum_{j=1}^L \left(
3X_j^{11}+2X_j^{22}+2X_j^{33}+2X_j^{44} \right.\no\\
&\phantom{=}&\left.
+X_j^{55}+X_j^{66}+X_j^{77}+X_j^{88}+X_j^{99}+X
_j^{10\,10}+X_j^{11\,11}
\right) 
\ee

These models with rung-mediated chemical potentials 
can again be solved via nested Bethe Ansatz.
For example, for the 3-tube we find 
\be
E_{3}^{\rm tube}(J)=L(1-3J)-\sum_{j=1}^{M_1} \left(
\frac 1{\left(\lambda_j^{(1)}\right)^2+\frac14} - 3 J \right)
\ee
where the roots $\lambda_j^{(1)}$ are given by the Bethe equations 
(\ref{nba}) with $N=8$.

Our results can be extended in a number of directions. 
Wang also introduced a solvable 2-leg ladder based on the 
supersymmetric permutation operator. The family of supersymmetric models 
will clearly lead to other solvable $n$-leg ladders and $n$-tubes.
Both of the solvable 2-leg ladders found by Wang have interesting physics,
with a transition to a rung-dimerised phase with a spin gap \cite{W}.
The physics of the solvable models presented here is expected
to be of equal interest, about which we hope to report in the
near future.  

It is a pleasure to thank Jon Links and Jaan Oitmaa for some helpful remarks.
This work is supported by the Australian Research Council.


\begin{thebibliography}{20}

\bibitem{DR}Dagotto E and Rice T M 1996 Surprises on the way from one- to 
two-dimensional quantum mag\-nets: the ladder materials {\it Science} {\bf 271} 618
%
\bibitem{W} Wang Y, Exact solution of a spin ladder model, cond-mat/9901168
%
\bibitem{AF} Albeverio S and Fei S-M, Exactly solvable models of
generalized spin ladders, cond-mat/9807341
%
\bibitem{KM} Kolezhuk A K and Mikeska H-J, Finitely correlated
generalized spin ladders, cond-mat/9803176
%
\bibitem{BY} See, for example, Batchelor M T and Yung C M 1995,
Integrable $su(2)$-invariant spin chains and the Haldane conjecture,
in {\em Confronting the Infinite},
ed A.~L. Carey et al (Singapore, World Scientific) p 167,
cond-mat/9406072,
and references therein.
%
\bibitem{S} Sutherland B 1975 Model for a multicomponent quantum system 
{\it Phys. Rev.} B {\bf 12} 3795

\end{thebibliography}
\end{document}